\documentclass[manuscript]{aastex61}

\accepted{July 6, 2018}

\shorttitle{Eruption on 2015 November 4}
\shortauthors{Wang et al.}

\begin{document}

\title{A solar eruption with relatively strong geo-effectiveness originating from active region peripheral diffusive polarities}

\correspondingauthor{Ying D. Liu}
\email{liuxying@swl.ac.cn}

\author{Rui Wang}
\affil{State Key Laboratory of Space Weather, National Space Science Center, Chinese Academy of Sciences, Beijing, China}
\affil{W.W. Hansen Experimental Physics Laboratory, Stanford University, Stanford, CA, USA}
\author{Ying D. Liu}
\affiliation{State Key Laboratory of Space Weather, National Space Science Center, Chinese Academy of Sciences, Beijing, China}
\affil{University of Chinese Academy of Sciences, Beijing 100049, China}

\author{Huidong Hu}
\affil{State Key Laboratory of Space Weather, National Space Science Center, Chinese Academy of Sciences, Beijing, China}
\affiliation{University of Chinese Academy of Sciences, Beijing 100049, China}

\author{Xiaowei Zhao}
\affil{State Key Laboratory of Space Weather, National Space Science Center, Chinese Academy of Sciences, Beijing, China}
\affiliation{University of Chinese Academy of Sciences, Beijing 100049, China}

\begin{abstract}
We report the observations of a moderate but relatively intense geo-effective solar eruption on 2015 November 4 from the peripheral diffusive polarities of active region 12443. We use space-borne Solar Dynamics Observatory and ACE observations. EUV images identified helical pattern along a filament channel and we regard this channel as flux-rope structure. Flow velocity derived from tracked magnetograms infers converging motion along the polarity inversion line beneath the filament channel. An associated magnetic cancellation process was detected in the converging region. Further, the pre-eruptive EUV brightening was observed in the converging region, the most intense part of which appeared in the magnetic cancellation region. These observations imply that the converging and cancelling flux probably contributed to the formation of the helical magnetic fields associated with the flux rope. A filament-height estimation method suggests that the middle part of the filament probably lies at a low altitude and was consistent with the initial place of the eruption. A thick current channel associated with the flux rope is also determined. For a expanding thick current channel, the critical height of the decay index for torus instability lies in the range of 37 -- 47 Mm. Southward magnetic fields in the sheath and the ejecta induced a geomagnetic storm with a $D_{st}$ global minimum of $\sim$ --90 nT.
\end{abstract}


\section{Introduction}
A solar eruption from 2015 November 4 gave rise to a geomagnetic storm with $D_{st}$ minimum of --90 nT, which disrupted satellite-based communications, including radar and GPS systems, and caused the air traffic grounded\footnote{\url{https://stillnessinthestorm.com/2015/11/earth-directed-solar-flare-and-cme/}}. Meanwhile, the associated transformer fires caused power outages and traffic issues. This is a lower-level eruption but produced obvious space weather effects. People are commonly willing to focus on the eruptions originating from the regions nearby main polarity inversion lines (PILs) of active regions (ARs) with strong magnetic fields. For instance, AR 11158 \citep{2012Sun,2013Rui,2015Vemareddy}, AR 11429 \citep{2013Liuxying,2013Simoes,2014Rui}, AR 11520 \citep{2014Cheng,2014Dudik,2016aRui,2016Hu}, and AR 12192 \citep{2015Sun,2016Jiang} all have strong magnetic shear or flux emergence. However, the present geo-effective eruption started from the peripheral diffusive polarities of an AR in a decay phase with relatively weak magnetic shear and flux emergence, which makes us interested in the physical mechanism of the formation of the eruption. Although the geo-effectiveness of the eruption from such a region is generally not as remarkable as from the regions around PILs of strong active regions, the investigation of the physical mechanism of the formation of such an eruption still has a certain significance for reducing the omissive predictions of harmful space weather events.

Photospheric motions play an important role in the formation and eruption of flux ropes. A converging motion as a prime photospheric motion was frequently mentioned in previous studies. A pair of magnetic field lines in a sheared core fields rooted on both sides of a neutral line can be pushed against each other by converging motions. Sequentially, a tether-cutting reconnection occurs at the intersection of the crossed arms of the sheared magnetic fields \citep{2001Moore,2017Vemareddy}. \citet{1989Vanballegooijen} found that converging flows can increase magnetic shear and promote the formation of helical flux-rope structures. In addition, converging motions can increase the upward magnetic pressure and cause flux ropes to deviate from an equilibrium state. Analytically, \citet{1995Forbes} found that even without flux cancellation a flux rope system subject to a converging motion would also experience a catastrophic behavior. A converging motion is imposed at the bottom boundary in a three dimensional magnetohydrodynamics (MHD) numerical simulations by \citet{2003Amari}, the results of which show that the flux rope always goes up.

AR background magnetic fields (or envelop fields) usually play an important role in determining the eruption of a flux rope (e.g., \citealt{2016bRui}). \citet{2005Torok}, \citet{2006Kliem}, \citet{2007Fan}, and \citet{2010Olmedo} investigated a torus instability of flux-rope structures, which indicates that the instability occurs when the Sun-directed Lorentz force due to the horizontal component of the background field decreases faster with increasing heights than the radial outward-directed ``hoop force.'' They used a decay index to measure the instability. \citet{1978Bateman} and \citet{2006Kliem} showed that the instability occurs when the critical value of the decay index n $\ge$ n$_{crit}$ = 1.5. The exact critical value of the decay index n$_{crit}$, at which the loss of equilibrium occurs, depends on the thickness and shape of the current channel associated with flux ropes \citep{2010Demoulin,2014Filippov,2015Zuccarello}.
The current channel is divided into thin and thick ones by the thickness, and into straight and circular ones by the shape. More precisely, the typical thickness, $2a$, must be small compared to the spatial scales of background magnetic fields and to the local radius $h$ of curvature of the current channel axis, i.e., $a \ll h$ for a thin current channel. For a thin current channel, previous analysis suggested that critical decay indices of 1 and 1.5 for straight and circular current channels, respectively. \citet{2010Demoulin} found that, for a thick current, if the current channel expands during an upward perturbation the critical decay index lies in the range n$_{crit}$= 1.1 -- 1.3 for both the straight and circular current channels.

Analysis of in situ measurements near the Earth is also performed. A flux-rope structure is reconstructed from the in situ measurements, and related features are compared with those of the flux rope in the solar coronal. Commonly, it is thought that few eruptions can occur in such diffusive polarities of the periphery of an AR. However, an eruption occurred in such a region but also gave rise to relatively strong geo-effectiveness, which is worth a thorough investigation. In this paper, Section 2 presents the observations on the Sun and the associated calculation results. Then the associated observation characteristics near the Earth are give in Section 3. Finally in Section 4, we give our conclusions and discussions.

\section{Observations on the Sun}
\begin{figure}

\epsscale{1}
\plotone{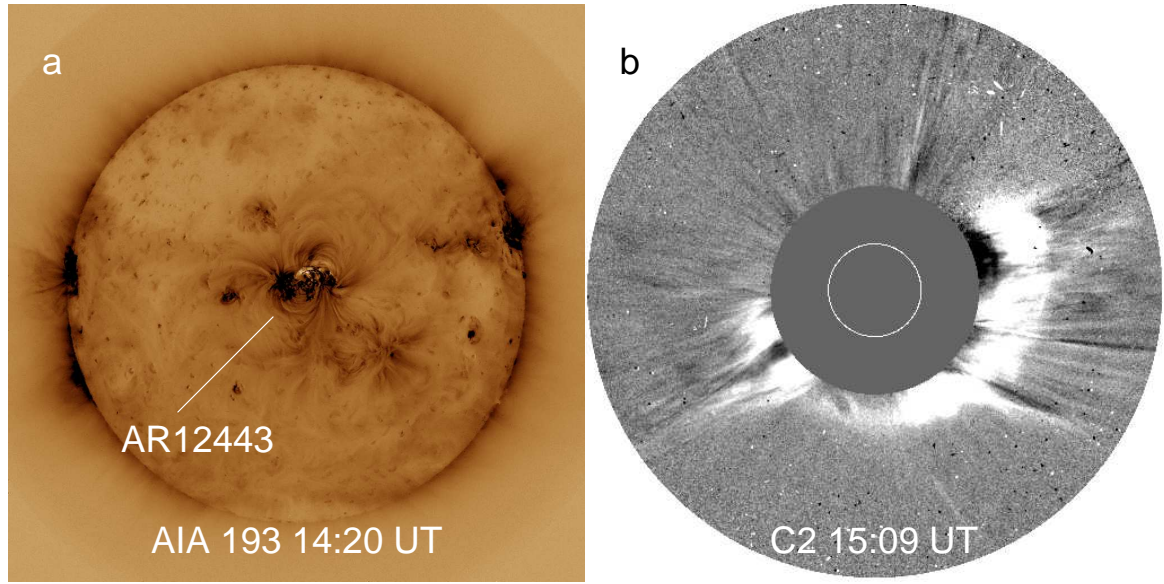}
\caption{Observations of the post-eruption and the initial phase of the CME. (a) shows the base-difference map of AR 12443 after the eruption at AIA 193 \AA~. The base image is at 13:30 UT. (b) shows the base-difference map of the halo CME in LASCO C2 view. The base image is at 13:26 UT. The white circle marks the location and size of the solar disk.\label{1}}
\end{figure}

NOAA AR 12443 (N06W10) was the only major AR on the solar disk (Figure \ref{1}a) on 2015 November 4, and it produced an M3.7 flare. The flare started at 13:31 UT, peaked at 13:52 UT, and ended at 14:13 UT. Soon after, a halo CME started emerging from LASCO C2 at 14:48 UT as shown in Figure \ref{1}b. The CME has many classical on-disk signatures. The GOES light curve indicates that the flare is a long duration event with extensive post-event loops visible in the EUV waveband. Large, fast wave emanated from the flare site with assocated dimming to the west and north of the AR. We obtain the CME linear speed using a graduated cylindrical shell model \citep{2006Thernisien} based on the observations from LASCO C2/C3 (no STEREO observations during this period). The linear speed within 15 $R_\odot$ is $\sim$565 $km~s^{-1}$.

\subsection{A flux-rope structure in a periphery of the AR}
\begin{figure}

\epsscale{1}
\plotone{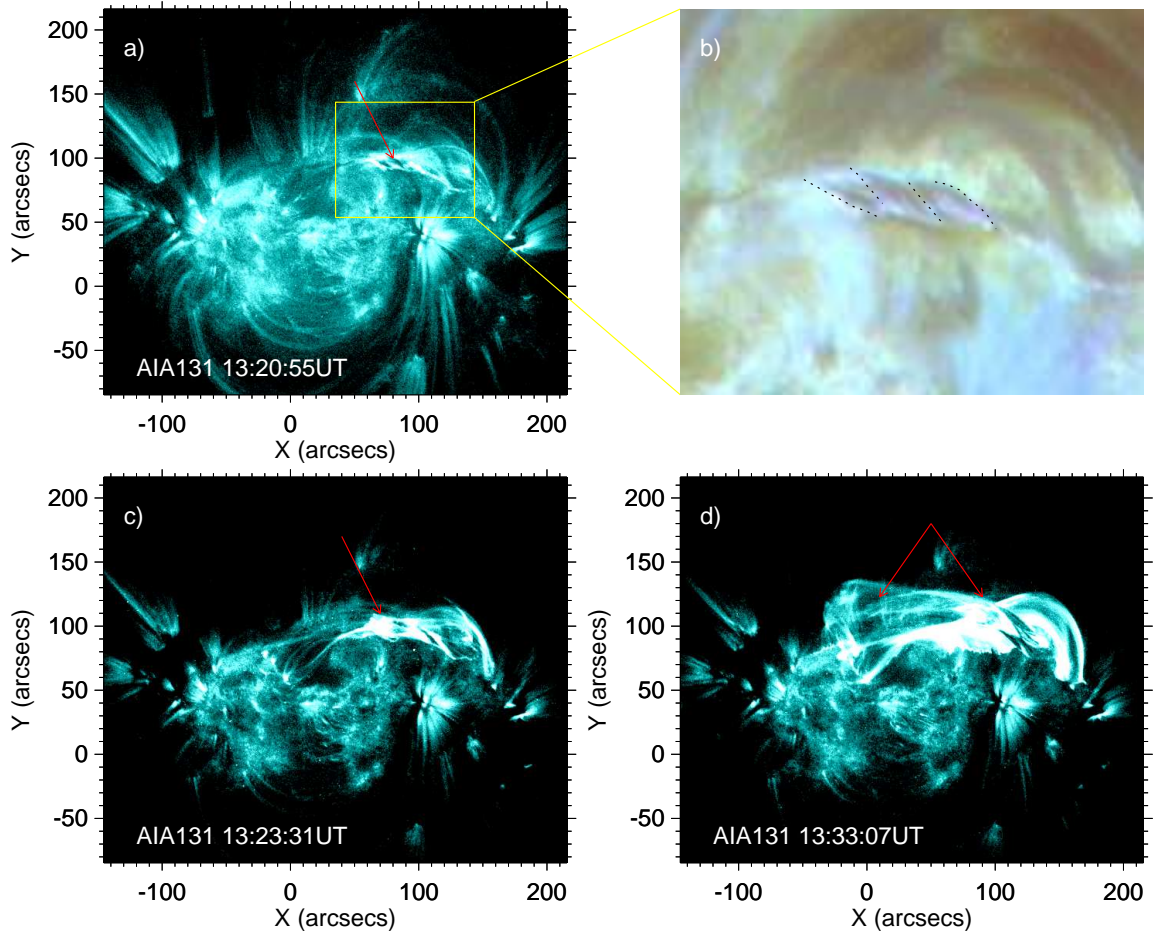}
\caption{The evolution of the flux rope at AIA 131 \AA~at the onset of the eruption in AR 12443. (a), (c), and (d) show that the flux rope as a hot channel was rising gradually. The red arrows indicate the flux rope. (b) shows the blowup of the flux rope in the AIA 211-193-171 composite image. The dotted lines mark the helical threads of the flux rope. \label{2}}
\end{figure}

A flux rope is a set of magnetic field lines wound around a central axis. It is also observed as an internal helical structure inside CMEs or ICMEs. On the Sun, it can be traced by a filament, as cold dense filament material can be collected in the lower parts of helical flux tubes, or observed as an evolving hot channel, the two ends of which anchor onto the photosphere (e.g., \citealp{2012Zhang}). The Atmospheric Imaging Assembly (AIA; \citealt{2012Lemen}) on board the Solar Dynamics Observatory (SDO; \citealt{2012Pesnell}) provided high temporal resolution full disk multi-thermal EUV images with a cadence of 12 s and 0$^"$.6 original pixel size. Figure \ref{2} shows the eruption processes in the CME source region at AIA 131 \AA~waveband. Figure \ref{2}a shows that a filament lay horizontally in the north of AR 12443. It seemed close to the photosphere at 13:20 UT before the onset of the M3.7 flare and only the western part can be observed at this time. A helical structure can be distinguished by the helical threads (see Figure \ref{2}a and the blowup of the structure in Figure \ref{2}b), and the first rising part corresponds to the helical threads (indicated by the red arrow in Figure \ref{2}c). Only 131 \AA~waveband can capture the overall arcade structure between two ends in Figure \ref{2}d, while other wavebands just exhibit the western part of the rising structure. We think that the rising helical structure is a flux rope associated with the filament.

The Helioseismic and Magnetic Imager (HMI; \citealt{2012Schou}) on board SDO provided high temporal resolution full disk line of sight (LOS) magnetic field data with a 45 s cadence and 0$^"$.5 original pixel size and photospheric vector magnetograms from the data product, called Space-weather HMI Active Region Patches (SHARPs; \citealt{2014Bobra,2014Hoeksema}) for AR 12443. We can see that the two ends of the flux rope are located at opposite sunspot polarities. The direction of the flux rope axial field is from east to west. AR 12443 was in a decay phase (see the unsigned flux curve in Figure \ref{3}b), so the sunspot group was disperse. The flux rope did not lie along the PIL of the strong magnetic field region in the center of the AR, but lay above the PIL of the relatively weak magnetic field region at the northern periphery of the AR (see the blue PIL in Figure 3a). From the side of the positive background magnetic polarity (on the north of the flux rope), the helical threads of the flux rope (marked by the black dotted lines in Figure \ref{2}b) are right bearing (the threads inclined from top left to bottom right). According to the method that \citet{1998Martin} used the filament barbs direction to judge the chirality of filaments, the flux rope is dextral (right-handed), but here directly using the helical pattern to judge instead of the barbs direction as a proxy.

For obtaining magnetic topology of the flux rope, we used two different kinds of non-linear force-free field (NLFFF) extrapolation methods, i.e., the optimization method \citep{2004Wiegelmann,2012Wiegelmann} and the GPU-DBIE method \citep{2000Yan,2006Yan,2013Rui} to reconstruct the flux rope, but we failed, i.e., we did not get a helical rope structure as we observed. Also, we adopted a forced extrapolation method \citep{2013Zhu,2016Zhu} which is based on the MHD model, and failed again. We analyzed why we failed and the reasons should be as follows.

(1) There was no strong shear detected at this period around the PIL. The reason could be that the observing instrument was unable to detect enough linear polarization signal from the anchoring points of the flux rope. Mathematically, the extrapolation methods based on a static boundary condition prefer a sheared boundary since shear on the boundary corresponds to twist in the corona.

(2) The strong negative sunspot polarity nearby interferes the magnetic field line tracing in such a weak field region. When the field line tracing step is not set small enough, the field line continuity is easily be broken at the high field gradient place.

(3) The AR was at its decay phase, the photospheric magnetic field might be compact and sheared several days ago or longer and the flux rope formed, but at this period the magnetic fields around the PIL were weak and not sheared enough. Consequently, the reconstruction is likely to affected by the reason (2).

(4)The flux rope maybe formed in a short time before the eruption, which was also proposed by \citet{2017Wang}. Note that the helical pattern of the flux rope in the EUV images was only observed within 30 minutes before the eruption.

\subsection{Influence of photospheric magnetic fields on the formation of the flux rope}
\begin{figure}

\epsscale{1}
\plotone{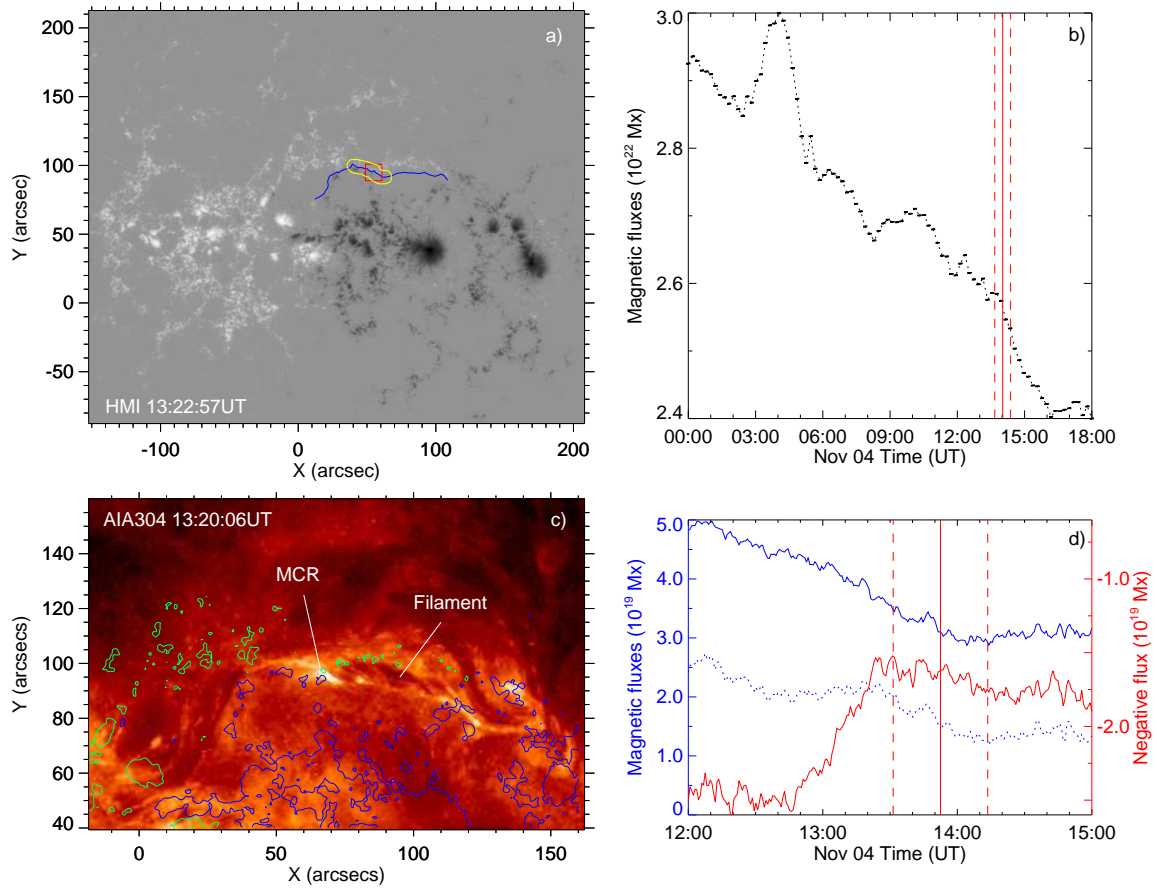}
\caption{(a) HMI LOS magnetogram for AR 12443. The small red box indicates the location of the magnetic cancellation region. The yellow contour is the converging region. The blue line marks the PIL beneath the filament. (b) Unsigned flux with error bars obtained over the entire AR by SHARP magnetograms. (c) Filament at AIA 304 \AA~wavelength in the pre-eruptive phase. The positive and negative polarities are presented by green and blue contours, respectively. (d) Signed magnetic flux of the positive (blue solid) and negative (red solid) polarities in the red box of (a). The integrated signed flux is presented by the blue dotted line. The red vertical dashed lines indicate the interval of the flare, and the peak time is marked by the vertical red solid line.\label{3}}
\end{figure}

By carefully checking the movie of the photospheric magnetic field evolution, we found that a small-scale magnetic cancelling feature appeared at the PIL of the periphery of the AR (small red box in Figure \ref{3}a). Figure \ref{3}c shows the low-lying filament at AIA 304 \AA~waveband before the eruption. The magnetic cancellation region (MCR; in the red box) happens to be exactly consistent with the pre-eruptive EUV brightening feature. The small red box shows the region where we used to calculate the evolution of the signed magnetic flux by LOS magnetograms. As shown in Figure \ref{3}d, the positive flux continues to decrease from 12:00 UT, and the negative flux starts to decrease around 12:45 UT (the start of obvious flux cancellation). We note that the negative flux stops decreasing when the eruption starts, which indicates that there is a relation between the magnetic cancellation and the eruption.

\begin{figure}

\epsscale{0.8}
\plotone{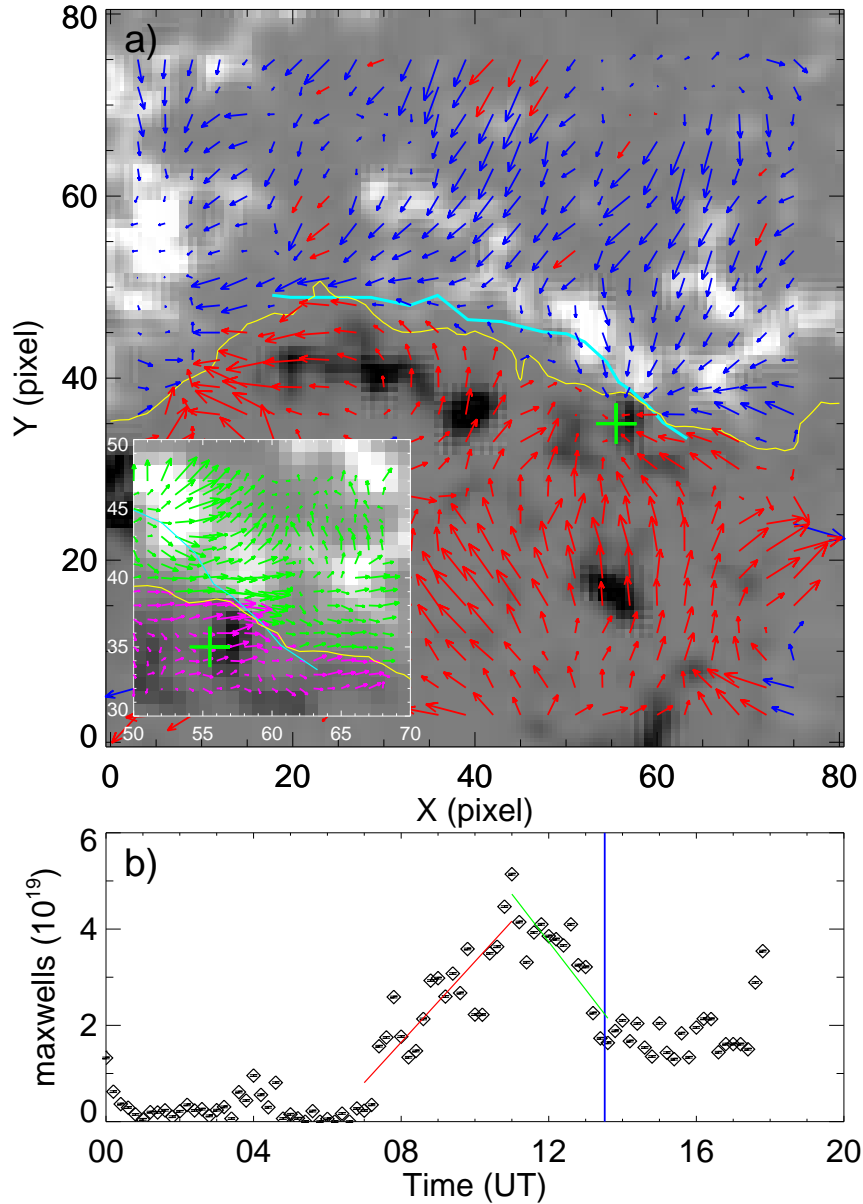}
\caption{(a) Horizontal velocity of flux motions derived from the DAVE4VM technique at 10:48 UT around the converging region in Figure \ref{3}a. The background is the B$_z$ map overplotted by horizontal velocities (blue/red arrows) in the positive/negative polarity. The yellow line represents the PIL. The cyan line is the projection of the EUV brightening in Figure \ref{3}c. The green cross marks the rough position of the magnetic cancellation region. The blowup in the lower left corner presents the transverse magnetic field vectors around the magnetic cancellation region. The fields (green/purple arrows) are in the positive/negative polarity. (b) Unsigned magnetic flux within the yellow contour in Figure \ref{3}a. The red and green lines are the least-square lines. The vertical blue line denotes the onset time of the eruption. The uncertainties are also plotted.\label{4}}
\end{figure}

We carefully checked the region around the MCR and found converging motions along the PIL. In Figure \ref{4}a, we plotted the velocity fields derived from DAVE4VM \citep{2008Schuck} on the B$_z$ map at 10:48 UT. The converging motions are along the yellow PIL and especially obvious in the MCR (around the green cross). Meanwhile, we projected the threadlike EUV brightening in Figure \ref{3}c as a cyan line onto this map. It shows that the EUV brightening is almost along the PIL. The intense part of the EUV brightening is just at the MCR. The pre-eruptive EUV brightening is generally thought to be the evidence of the magnetic reconnection associated with the eruption. Figure \ref{4}b shows the unsigned magnetic flux within the yellow contour of Figure \ref{3}a calculated with the SHARP data. The horizontal component B$_h$ of the photospheric magnetic fields along the PIL of the periphery of the AR is relatively weak. When we set B$_h$ above a threshold of 100 G, few pixels are left. It implies that the uncertainties of B$_h$ are large. By contrast, the analysis to the vertical component B$_z$ is more reliable. The pixels that contribute to the unsigned magnetic flux calculation are selected by examining two data segment maps in the SHARP files: BITMAP and CONF\_DISAMBIG. The BITMAP segment identifies the pixels located within the bounding curve (BITMAP $\ge$ 30). The CONF$\_$DSIAMBIG segment identifies the final disambiguation solution for each pixel with a value which maps to a confidence level in the result (disambiguation noise threshold $\approx$ 150 G, CONF\_DISAMBIG = 90; \citealt{2014Bobra}).

\begin{figure}

\epsscale{1}
\plotone{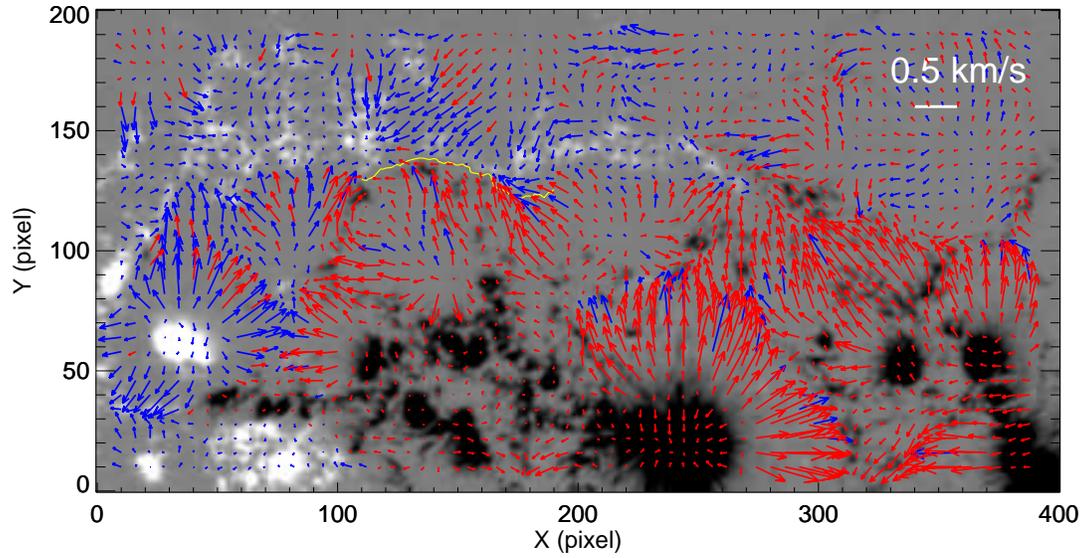}
\caption{Flow map averaged from 07:00 UT to 12:00 UT. The background is the B$_z$ map overplotted by horizontal velocities (blue/red arrows) in the positive/negative polarity. The yellow line represents the PIL in the converging region. Flow speed of 0.5 $km~s^{-1}$ is denoted by the short white line.\label{converging}}
\end{figure}

The unsigned magnetic flux in Figure \ref{4}b increases from 07:00 UT to 12:00 UT and decreases from 12:00 UT to the beginning of the eruption. The increase is due to the converging motion of the photospheric magnetic fields mentioned above, i.e., magnetic fields continuously flow towards the PIL from both sides and converge into the yellow contour. Note that when the prime converging flows from both sides of the PIL were injected perpendicular to the middle part of the PIL (see Figure \ref{4}a), antiparallel motions along the PIL formed on both sides of the prime converging flows, which could be treated as shearing motions. The decrease of the flux should correspond to the magnetic cancellation shown in Figure \ref{3}d. The blowup in Figure \ref{4}a shows that the transverse magnetic fields are parallel to the PIL in the MCR, which implies that the magnetic fields along the PIL are sheared. \citet{1989Vanballegooijen} found that shearing and converging motions, by the reconnection of sheared magnetic fields, can result in the formation of long helical loops and the submergence of short loops, which is observed as flux cancellation \citep{1984Rabin}. Figure \ref{converging} shows the flow map averaged from 07:00 UT to 12:00 UT, which presents obvious converging motions within the yellow contour of Figure \ref{3}a. We note that the helical pattern in Figure \ref{2}b just formed in the converging region. Also, the rising part of the flux rope initiated from this region (see Figure \ref{2}c). Therefore, we think that the converging motions probably play a role in promoting the formation of the helical magnetic field associated with the flux rope.

A eruption can occur often due to the loss of equilibrium of the magnetic system. \citet{2017Wang} measured a very strong twist of the flux rope during the eruption. Kink instability probably occurred at the initial stage of the eruption. However, kink instability is not an effective mechanism for full solar eruptions. Hardly any existing kink model alone successfully produces a CME. It often needs to cooperate with a torus instability (e.g., \citealp{2006Kliem,2014Vemareddy}). The eruption occurred at the edge of the AR. The constraining force of the background fields should be weaker than the magnetic center.

\subsection{Influence of background magnetic fields on the eruption}
\begin{figure}

\epsscale{0.8}
\plotone{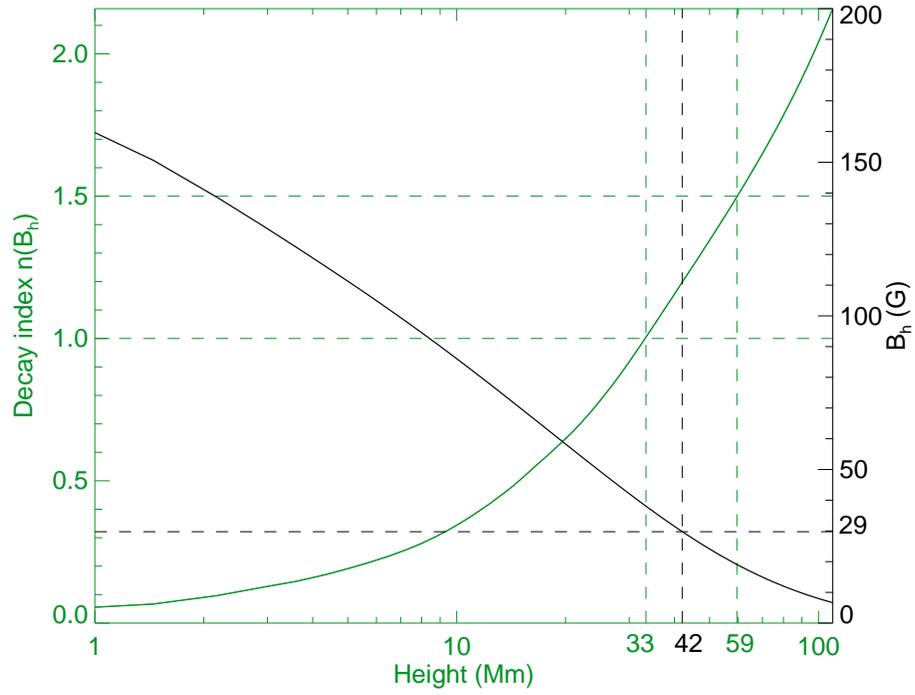}
\caption{Height profile of mean $B_h$ (black) and average decay index $\langle n\rangle$ (green) above the PIL in the periphery of the AR. The critical values n$_{crit}$ = 1 and 1.5 correspond to the heights of 33 Mm and 59 Mm, respectively. The mean $B_h$ for a typical height of eruption onset of 42 Mm is only 29 G.\label{5}}
\end{figure}

As discussed above, we checked the magnetic strength of the overlying magnetic fields and the torus instability of the flux rope. Torus instability was measured by a decay index as a function of height. Figure \ref{5} shows the height profile of $B_h$ (black curve) and the average decay index $\langle n\rangle$ (green curve) above the PIL of the periphery of the AR
\begin{equation}
n=\frac{\partial{\ln B_h}}{\partial{\ln h}},
\end{equation}
where $B_h$ is the horizontal magnetic field component and $h$ is the height above the photosphere. For comparison, we refer to the results of Table 1 in \citet{2015Sun} and show them with our results in Table 1. The critical height of AR 12443 is only lower than AR 12192, which produced a confined eruption, and higher than the other two ARs that are both eruptive. Moreover, all of the first three ARs produced X-class flares. For the critical height of AR 12443, it seems that there is no better condition provided for the eruption. That is also revealed through $B_h$(42)/$B_h$(2). However, we would like to mention that the parameter $B_h$(42) of AR 12443 is the lowest among these ARs. $B_h$(42) indicates the mean horizontal field component $B_h$ at 42 Mm, a typical height of eruption onset \citep{2008Liu}. The lower value of $B_h$(42) in AR 12443 implies that there were weaker background magnetic fields and weaker constraining force over the flux rope. Therefore, more detailed analysis is needed on the decay index distributions at different heights, since the relative positions between flux ropes and decay index distributions in space can sometimes exhibit different results from the average decay index \citep{2016bRui,2017Chandra}.

\begin{figure}

\epsscale{1}
\plotone{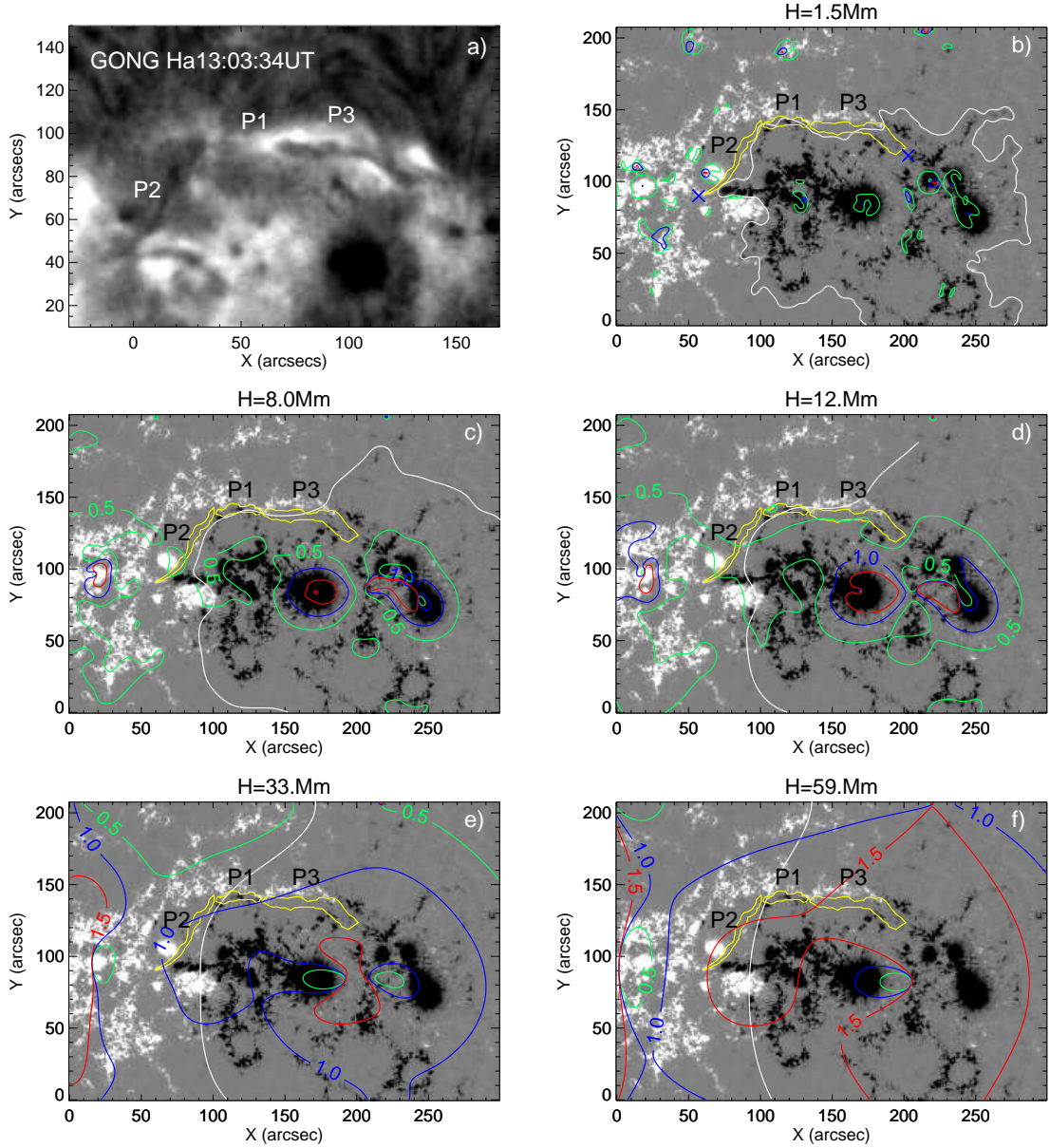}
\caption{(a) GONG H$_\alpha$ image above the PIL in the periphery of the AR. (b)-(f) Distributions of the decay index n at different heights in Mm on the SHARP B$_z$ map. The green, blue, and red contours correspond to the decay index of 0.5, 1, and 1.5, respectively. The white lines denote the PIL of the magnetic field by potential field extrapolation at the heights of 1.5, 8, 12, 33, and 59 Mm, respectively. The yellow contours are the filament components extracted from (a) and projected from GONG image onto the SHARP map. The possible locations of the ends of the filament are labeled by blue crosses in (b).\label{6}}
\end{figure}

We calculated the distributions of decay index values at different altitudes. Figure \ref{6}a shows the H$_\alpha$ image from the Global Oscillation Network Group (GONG) program. A filament structure was lying above the PIL. Although it seems that there are two separated parts of the filament, they are just two parts of an integrated filament before the EUV brightening (also see the filament in Figure \ref{3}c). The middle region (P1) became less visible in the H$_\alpha$ wavelength probably for two reasons. One is the heating of the filament which can suppress the absorption of H$_\alpha$ radiation. The magnetic reconnection at a low height can make such heating. The other is the Doppler shift in the moving material which can remove its H$_\alpha$ line out of the filter passband. A reasonable explanation is that, after the appearance of the EUV brightening (magnetic reconnection occurred), this cancelling region was in a dynamic state and ready to get into an erupting stage.

Due to the correlations between filaments and magnetic flux ropes, we can trace the flux rope by the filament at the initial stage of the eruption. Unfortunately, we cannot measure the height of the filament directly since STEREO is behind the Sun and cannot provide efficient two view-point observations of all the eruptions throughout the November. Fortunately, according to the method of \citet{2016bFilippov}, we can estimate the height of filaments. It is based on the confirmed by-observations assumption that the material of filaments is accumulated near coronal magnetic neutral surfaces \citep{2016aFilippov}. Therefore, a PIL at each height is a favorable place for horizontal equilibrium of a stable flux rope, but it is only a necessary condition. Another necessary condition for the stable equilibrium is the quantity of the decay index below the critical value. In fact, flux ropes may stay only in few places where both conditions are fulfilled.

Figure \ref{6}b-\ref{6}f shows the relative positions of the filament (yellow), the PILs at different heights (white), and decay index iso-contours (green, blue, and red). The background in Figure \ref{6}b-\ref{6}f shows the B$_z$ magnetogram taken by the SHARP data at 13:00 UT on 2015 November 4, with a pixel size of 1$^"$. The overall extrapolation boundary is larger than the background and flux-balanced. The extrapolation was computed using a Fourier representation based on Green¡¯s function \citep{1978seehafer}. The filament contours were extracted from the GONG H$_\alpha$ image and projected onto the SHARP Cylindrical Equal-Area magnetograms. Also, considering the projection effects of the filament and the PILs changing with heights \citep{2016bFilippov}, we shifted the PIL for a given height $h$ by the horizontal coordinate $x$ and vertical coordinate $y$ on the map by values
\begin{eqnarray}
\Delta x &=& h\,tan{\lambda}_0, \\
\Delta y &=& h\,tan{\phi}_0,
\end{eqnarray}
where $\lambda_0$ and $\phi_0$ are longitude and latitude of the selected area center. According to the rules above, the middle part (P1) of the filament follows the PIL at 1.5 Mm, and is outside the 0.5 decay index iso-contour (green) which means the flux rope is in a region that the decay index value is less than 0.5. Therefore, 1.5 Mm is probably a favorable height for a stable P1. While for the eastern (P2) and western part (P3), only the places near the ends of the filament follow the PIL (see Figure \ref{6}b). At 8.0 Mm, the whole P2 does not follow the PIL. The cancelling region of P1 and most part of P3 follow the PIL. However, at 12 Mm, only the cancelling region follows the PIL. According to Figure \ref{6}b-\ref{6}d, we can envisage the configurations of the flux rope, i.e., two ends of the flux rope anchor on the positive and negative polarities, respectively (refer to the blue crosses in Figure \ref{6}b), the middle of the flux rope (P1) is close to the photosphere. P3 stays low but not as low as P1. P2 stays relatively higher. The overall flux rope looks like an asymmetric M-shaped structure. This is consistent with the observations that the helical patterns associated with the flux rope started emerging in the converging region (see Figure \ref{2}a), which is probably related to photospheric motions. Figure \ref{6}e and \ref{6}f show the decay index iso-contours of n = 1, 1.5, which correspond to the heights at 33 Mm and 59 Mm, respectively. Of particular interest is that at both altitudes the region of decay index above the critical value (n $\ge$ 1 or n $\ge$ 1.5) begins to cover the overall filament, which implies that both altitudes could become the critical height. According to previous studies, if thin current channels associated with a flux rope are considered, then n$_{crit}$ = 1 for a straight current channel, and n$_{crit}$ = 1.5 for a perfectly circular current channel. However, our estimation results show that the flux rope stays low, the thickness of the current channel is comparable with its height i.e., it should be regarded as a thick current channel. \citet{2010Demoulin} found that at the limit of thick current, if the current channel expands during an upward perturbation, the critical decay index lies in the range n$_{crit}$ = 1.1 -- 1.3 for both the straight and circular current channels. Figure \ref{2} shows that the hot channel is deformable and was expanding during the eruption. According to their findings, the critical height lies in the range h$_{crit}$ = 37 -- 47 Mm. Compared with the critical height of the other ARs in Table 1 again, AR 12443 has a comparable critical height with the strong AR 11429 with 34 Mm critical height which had two successive solar eruptions \citep{2013Liuxying,2014Rui}, and AR 11158 with 42 Mm critical height.

\section{In situ measurements near the Earth}

\begin{figure}

\epsscale{0.6}
\plotone{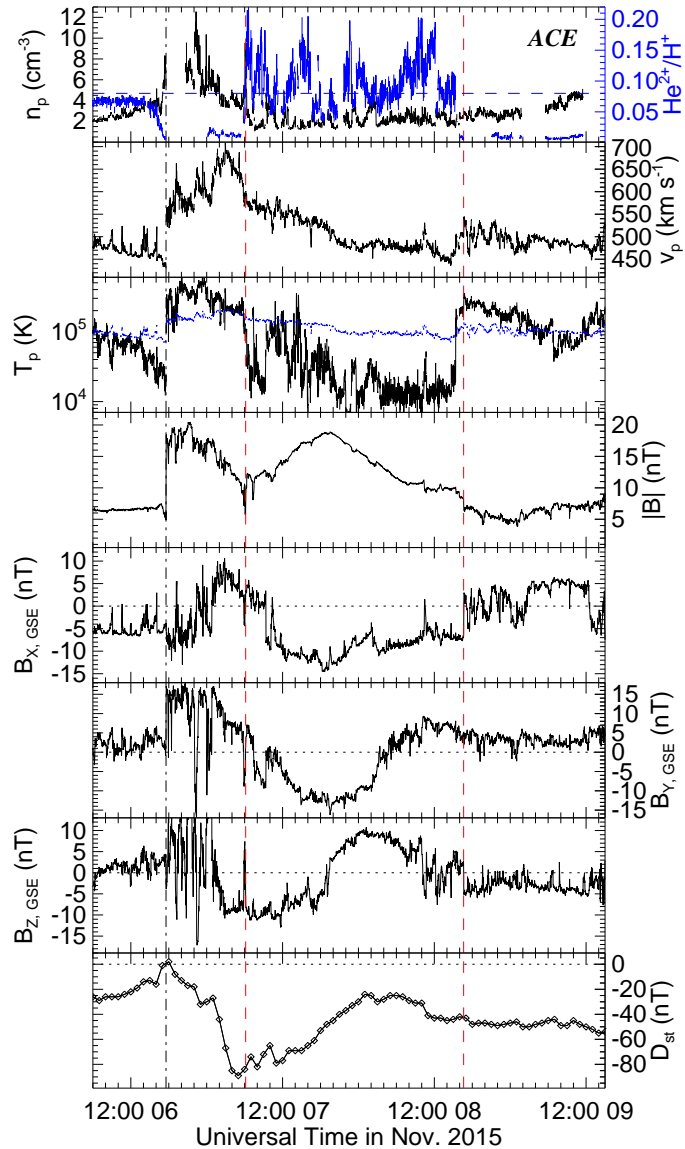}
\caption{Solar wind measurement near the Earth. From top to bottom, the panels show the proton density (with the alpha-to-proton density ratio overlaid in blue), bulk speed, proton temperature (overlaid with the expected proton temperature calculated from the observed speed \citep{1987Lopez}), magnetic field strength and components, and D$_{st}$, respectively. The D$_{st}$ profile is shifted 1 hr earlier for comparison with the in situ measurements. The black vertical line marks the arrival time of the shock, and the red vertical lines indicate the interval of the flux-rope structure determined by the GS reconstruction.\label{7}}
\end{figure}

Figure \ref{7} shows the associated in situ measurements at ACE during the event. A shock passed ACE at 17:34 UT (the vertical black dashed-dotted line) on November 6 and caused the sudden commencement of the geomagnetic activity. The alpha-to-proton density ratio (in the top panel) is above the threshold (horizontal blue dashed line) for a typical ICME (e.g., \citealp{2004Richardson,2005Liuxying}), and the low proton temperature \citep{1995Richardson} and the magnetic field profile can help us identify that a flux-rope structure between the vertical dashed lines in Figure \ref{7}. The $B_x$ component remains negative, while $B_y$ and $B_z$ components change signs from negative to positive inside the ICME. This means that the flux rope belongs to the south--west--north (SWN) category according to the classification schemes introduced by \citet{1998Bothmer} and \citet{1998Mulligan}. The $D_{st}$ profile indicates a geomagnetic storm sequence with a global minimum of $\sim$ --90 nT. The decrease of the $D_{st}$ is related to both the southward magnetic field components in the sheath region of the shock and in the flux rope. According to the arrival time (17:34 UT November 6) and eruption time (13:31 UT November 4), we can roughly estimate that the average propagation speed of the CME from the Sun to the Earth is $\sim$ 800 $km~s^{-1}$.

\begin{figure}

\epsscale{1}
\plotone{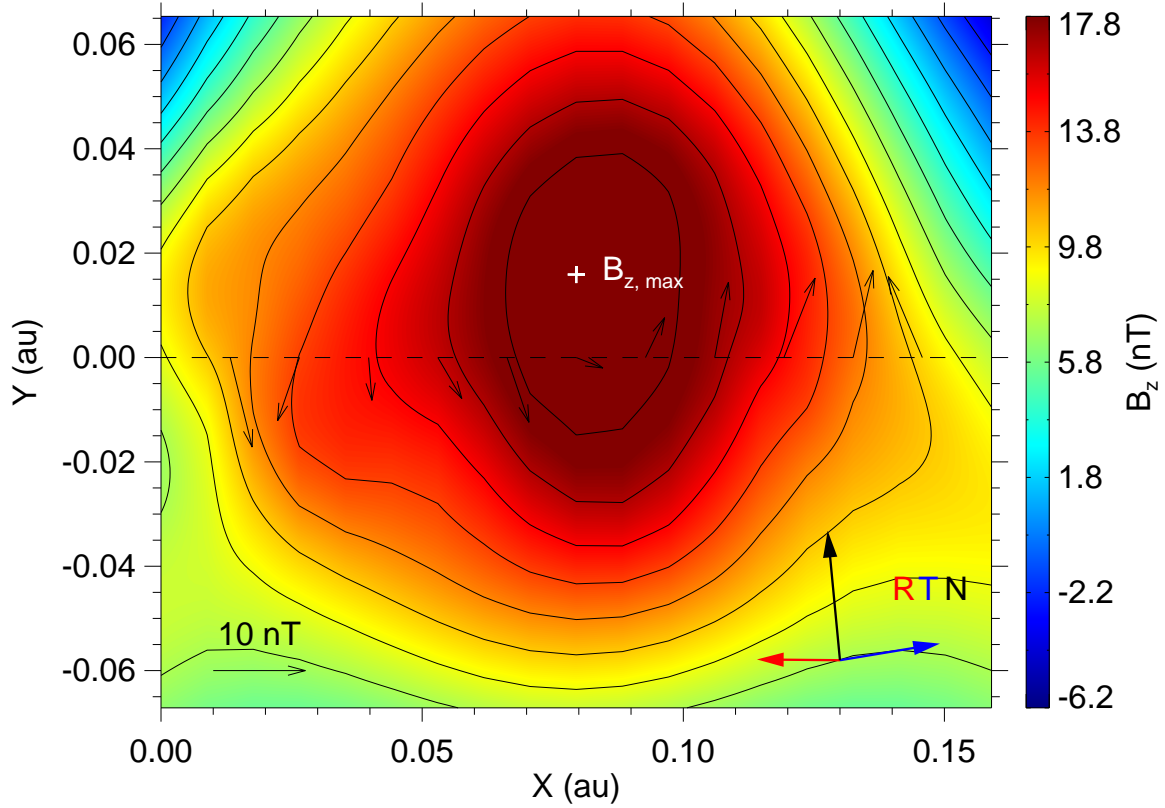}
\caption{Reconstructed cross section of the flux rope near the Earth. Black contours show the distribution of the magnetic flux function (\citealp{2002Hu}), and the color shading indicates the value of the axial magnetic field scaled by the color bar on the right. The location of the maximum axial field is indicated by the white cross. The dashed line marks the trajectory of the Earth. The thin black arrows denote the direction and magnitude of the observed magnetic fields projected onto the cross section, and the thick colored arrows show the projected RTN directions.\label{8}}
\end{figure}

Grad-Shafranov (GS) reconstruction method \citep{1999Hau,2002Hu} is sensitive to the chosen boundaries, which can help determine the flux-rope interval at Wind in combination with the plasma and magnetic field parameters and has been validated by well-separated multi-spacecraft measurements (e.g., \citealp{2008Liuxying}). Figure \ref{8} shows that the GS reconstruction gives a right-handed chirality flux-rope structure for the ICME, as can be determined from the transverse fields along the spacecraft trajectory together with the axial direction of the flux rope. The elevation angle and azimuthal angle were --5$^\circ$  and 39$^\circ$ in the RTN coordinates, respectively. The elevation angle is very small, which generally implies that the flux rope was lying in the ecliptic plane. The chirality and orientation of the flux rope were consistent with the observations on the Sun. For the small inclination angle, the geomagnetic activity was mainly caused by the azimuthal magnetic field components (about 10 nT) of the flux rope rather than the axial components. However, the maximum of the axial field $B_{axial}$ is stronger, $\sim$ 18 nT. That is to say, if the flux rope were not lying in the ecliptic plane but perpendicular to the ecliptic plane and southward inclined, the geomagnetic activity would be stronger than observed. This indicates the importance of forecasting the orientation as well as the field components of the flux-rope structure of an ICME (e.g., \citealp{2015Liuxying,2015Vemareddyb,2016Hu}).

\section{Conclusions and Discussions}

NOAA AR 12443 produced a CME eruption which gave rise to relatively strong geo-effectiveness. We consider the influences of photospheric motions and background magnetic fields on the eruption. The former is probably associated with the formation of a flux rope. The latter probably determines the final eruption. The main results of our study are described as follows.

Flow velocity derived from tracked vector magnetic field reveal converging motion of magnetic features along the PIL. The evolution of the unsigned magnetic flux shows decreasing trend confirming the converging and cancelling processes of the photospheric magnetic fields where a helical pattern related to the flux rope formed prior to the eruption. More precisely, the pre-eruptive EUV brightening is along the PIL of the converging region and its intense part overlapped the magnetic cancellation region. The continuous converging motions probably caused the magnetic reconnection along the PIL forming the helical magnetic fields associated with the flux rope.

Background magnetic fields play an important role in determining an eruption. The decay of the background fields is related to torus instability. The critical height when $\langle n\rangle$ = 1.5 is 59 Mm. Although 59 Mm as a critical height is not high, it is not better than those in the ARs with strong magnetic fields exhibited by \citet{2015Sun}. The analysis of the decay index distributions was then carried out at different altitudes. It shows that the decay index iso-contours of n = 1, 1.5 can both cover the overall flux rope, which implies that the corresponding height of 33 and 59 Mm can both become the critical height for the torus instability. According to previous analysis, the exact critical value of the decay index n$_{crit}$ depends on the relative thickness $2a$ and shape of the current channel associated with the flux rope. The height $h$ of the current channel determines its relative thickness, since $a \ll h$ for a thin current channel. A filament-height estimation method was adopted, which is by means of the relative distributions of the decay index and the PIL. It makes up the absence of multi-viewpoints observations from STEREO. The middle part of the flux rope was estimated to stay in a low altitude ($\sim$1.5 Mm). According to the analysis of \citet{2010Demoulin}, we regard it as a thick current channel. Their analysis also indicated that the critical decay index lies in the range n$_{crit}$ = 1.1 -- 1.3 for an expanding thick current channel. Therefore, the critical height should lie in the range h$_{crit}$ = 37 -- 47 Mm. AR 12443 has a comparable critical height with the ARs with strong magnetic fields in the study of \citet{2015Sun}.

The associated characteristics near the Earth are presented. According to the in situ observations and GS reconstruction, both the orientation and the chirality of the flux rope in the ICME agree with those on the Sun. Southward magnetic fields in the sheath and the ICME (azimuthal components $\sim$10 nT) are correlated with the enhancement of the geomagnetic activity. If the flux rope were perpendicular to the ecliptic plane, the stronger axial field component ($\sim$18 nT) would cause a stronger geomagnetic activity. This indicates the importance of forecasting both the orientation and field components of the flux rope in an ICME from the morphological characteristics of a flux rope on the solar surface (e.g., \citealp{2015Liuxying,2016Hu}).

In this work, magnetic polarities in the periphery of an active region stayed in a decay phase, which implies that the magnetic non-potentiality of the photospheric fields is weakening. The eruption with relatively strong geo-effectiveness still can be triggered in such a region. Therefore, understanding the physical mechanism of the eruption from such a region is significant for space weather forecasting.

\acknowledgments
We are grateful to the anonymous referee for the extensive and constructive comments and suggestions that helped us to improve the quality of the manuscript. The work was supported by NSFC under grants 41604146, 41774179, and 41374173, the Specialized Research Fund for State Key Laboratories of China, and Strategic Priority Program on Space Science (No. XDA 15011300). The authors gratefully acknowledge financial support from China Scholarship Council. The data used here are courtesy of the NASA/SDO HMI and AIA science teams. We thank the HMI science team for making the SHARPs vector magnetograms available to the solar community.

\begin{deluxetable*}{cccccc}[b!]
\tablecaption{Comparison of the background field of four ARs above the PIL\label{tab:mathmode}}
\tablecolumns{6}
\tablenum{1}
\tablewidth{0pt}
\tablehead{
\colhead{} &
\colhead{AR 12192\tablenotemark{a}} &
\colhead{AR 11429\tablenotemark{a}} & \colhead{AR 11158\tablenotemark{a}} & \colhead{AR 12443}  & \colhead{Unit}\\
}
\startdata
Critical height & 77 & 34 & 42 & 59 & Mm \\
B$_h$(42)/B$_h$(2) & 0.35 & 0.06 & 0.05 & 0.21 \\
B$_h$(42) & 220 & 61 & 42 & 29  & G \\
\enddata
\tablenotetext{a}{ The parameters of the three ARs come from the results of \citet{2015Sun}.}
\end{deluxetable*}

\end{document}